# Photo-induced Macro/Mesoscopic Scale Ion Segregation in Mixed-Halide Perovskites: Ring Structure and Ionic Plasma Oscillations

Xiaoxiao Sun[1,2,*], Yong Zhang[3,*], Weikun Ge[4]

[1] Laboratory for Thin Films and Photovoltaics, Empa−Swiss Federal Laboratories for Materials Science and Technology, 8600 Duebendorf, Switzerland

[2] Department of Information Technology and Electrical Engineering, ETH Zurich, 8093 Zurich, Switzerland

[3] Department of Electrical and Computer Engineering, The University of North Carolina at Charlotte, Charlotte, NC 28223, USA

[4] Department of Physics, Tsinghua University, Beijing 10084, People's Republic of China

**Abstract**

Contrary to the common belief that light-induced halide ion segregation in a mixed halide alloy occurs within the illuminated area, we find that the Br ions released by light diffuse away from the area, which generates a counter-balancing Coulombic force between the anion deficit and surplus region, together resulting in a macro/mesoscopic size anion ring surrounding the center, showing a photoluminescence ring. Upon removing the illumination, the displaced anions return to the illuminated area, and the restoring force leads to a damped ultra-low-frequency oscillatory ion motion, which may be the first observation of an ionic plasma oscillation in solids.

Lead halide perovskites (e.g., MAPbI$_3$) are an emerging family of semiconductor materials with excellent optoelectronic properties ideally suited for photovoltaic and light-emitting applications [1–3]. This class of soft crystals is known to be mixed conductors of both electronic and ionic conductivity. Significant ion migration has been reported in these materials and is one of the main mechanisms responsible for anomalous I-V hysteresis and poor stability in perovskite solar cells [4–6]. Particularly, as initially reported in 2015 [7], mixed halide perovskites (e.g., MAPbI$_{1-x}$Br$_x$) further exhibit photoinduced halide anion segregation [7–11]. It is generally asserted that a uniform alloy MAPbI$_{1-x}$Br$_x$ experiences anion segregation into Iodide-rich and Bromide-rich domains respectively under continuous above-bandgap illumination and the process is reversible when the illumination is removed. A number of microscopic mechanisms have been proposed to explain the phenomenon [12–16]. They can be summarized into three categories [17]: 1) Thermodynamic in origin where halide composition/bandgap differences favor halide demixing under illumination; 2) polaron-induced lattice strain; 3) photogenerated carriers gradient or electric fields, interacting with point defects. However, none of them can unambiguously explain all the key aspects of the phenomenon [17]. In fact, the chemical and structural nature of the so called "Br-rich" and "I-rich" regions are not yet well understood, although they are implicitly assumed to be simply Br-rich and I-rich alloys, respectively.

Note that previous studies were typically performed under uniform illumination over a macroscopic sample area, where the overall volume of the "I-rich" regions was found to increase with illumination time, eventually the illuminated area became almost "fully converted", if judging from photoluminescence (PL) spectra [18]. However, it has been noted that even in the fully converted state, the "I-rich" regions only represents a small fraction of the overall volume, based on the absorption strength and XRD intensity of the "I-rich" regions [7]. Regardless which model was invoked, it was always taken for granted that the segregation occurred within the illuminated area, and implicitly assumed that there was no change in cation-anion stoichiometry other than local exchange of the anions in the similar manner as in composition modulation in other semiconductor alloys [19]. If the Br ions remained in the illuminated area, as implied, the complete quenching of the alloy state PL would suggest that the "I-rich" regions were uniformly embedded within the "Br-rich" matrix with an average separation shorter than the carrier diffusion length of the "Br-rich" matrix. Since the carrier diffusion lengths in such polycrystalline films are relatively short, typically in the order of $\mu m$ [20] and comparable to the polycrystalline domain sizes, the average separation of the "I-rich" regions is expected to be rather small. It has been suggested that the sizes of the "I-rich" regions are

sub-10 nm [14,21]. In short, previous effort has focused on the microscopic scale anion segregation within the illuminated area under uniform light illumination.

In contrast, we discover that the anion segregation in the mixed halide alloys is a non-local effect of which the ion redistribution may occur in a macroscopic or mesoscopic scale, proportional to the illumination beam size, up to well over $mm$. Specifically, the Br ions are expelled from the illuminated area, resulting in a positively charged area of an off-stoichiometry alloy, and concurrently forming a negatively charged Br-rich ring. This phenomenon can be viewed as an ionic analogy of a mesoscopic PL ring formation away from the illuminated site in GaAs/AlGaAs QWs, resulting from the disparity in the electron and hole diffusion lengths and thus their spatial profiles [22–25]. Furthermore, we observe an ultra-low-frequency oscillatory motion of Br ions between the ring and center, an oscillation of ionic plasma, which has not been reported before in solids and no electronic equivalence. Besides their own significance, these new findings offer new insights to the underlying mechanism of the ion segregation in the mixed halide alloys.

Mixed-halide $MA_{0.17}FA_{0.83}Pb(I_{0.5}Br_{0.5})_3$ thin films on glass substrates are used (see Methods for details in Supplemental Material, SM). The sample is illuminated with a 639 nm laser beam, with the beam size varying from 1 $\mu m$ to 1 $mm$, to induce the halide ion redistribution. The dynamics of the ion motion is probed in-situ by spatially-resolved continuous-wave (CW) and time-resolved (TR) PL mapping and CW absorption as a function of time (up to 130 hours) over an area much beyond the illuminated area itself. All measurements are performed under ambient condition.

**Figure 1** shows time evolution of the light-induced spatial modification in PL that reflects the halide ion distributions over a mesoscopic scale about 10 times of the illumination beam size (~12 $\mu m$). The as-grown sample shows a PL peak at ~670 nm, typical of the alloy with $x_{Br} = 0.5$, and the 670 nm alloy peak decreases in intensity, accompanying by a redshift of the overall emission band to ~790 nm after prolonged illumination, as shown in **Fig. 1b**, which is consistent with the commonly observed light induced halide ion segregation process and the latter peak is attributed to the emission from "I-rich" regions [14,18,26–29]. As references, **Fig. 1a** shows the confocal PL mapping results at 670 and 790 nm (in 10 nm bandwidth), respectively, under a low excitation density of 0.3 W/cm$^2$. The sample is then illuminated at the center of the mapped area with the same beam size but a higher density of 1.5 W/cm$^2$. We further perform PL mapping for the 670 and 790 nm emissions, respectively, over the same area under the same conditions as in **Fig. 1a** at different illumination times. As shown in **Fig. 1c**, upon the illumination is turned on, a dark circle starts to form in the 670 nm PL mapping, while a complementary and comparable size bright circle appears in the 790 nm mapping, and the size and contrast of the two circles increase with time. More interestingly, in the meantime, a bright ring of the 670 nm emission emerges and establishes outside the illuminated area, resulting in a donut shape intensity distribution.

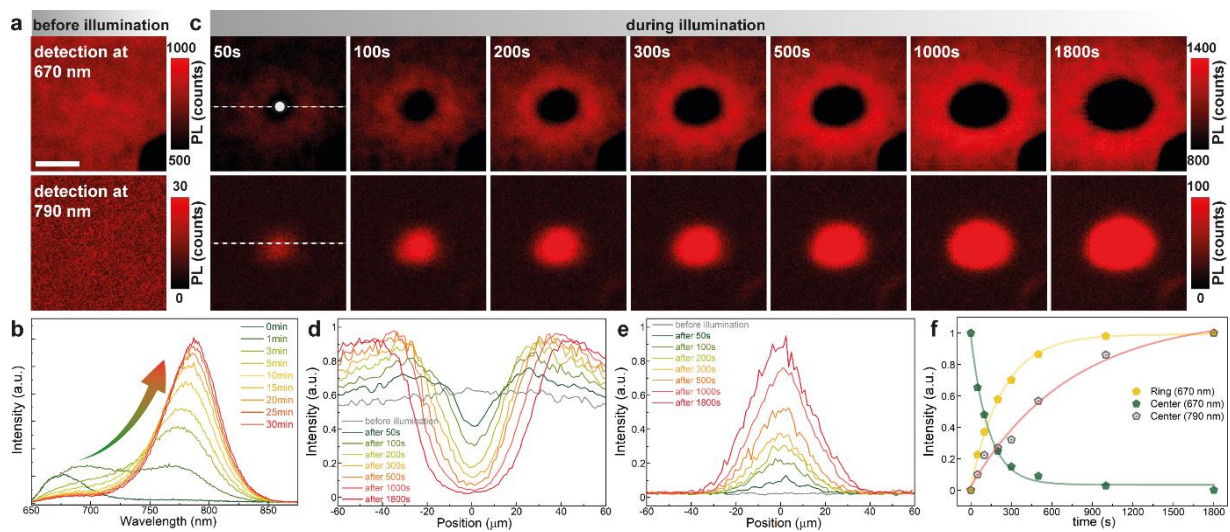

**Figure 1.** (a) PL mapping at 670 and 790 nm before illumination (The dark area at the lower right corner is due to a marker). The scale bar is 40 $\mu m$. (b) PL spectra from the illuminated site at the center of the mapped area in (a) over a time of 30 min. (c) PL mapping at 670 nm (top row) and at 790 nm (bottom row) of the same area as in (a) during light illumination. The small circle indicates the beam size. (d) & (e) Time dependent PL spatial profiles taken along the white dashed lines from PL maps for 670 and 790 nm in (c), respectively. (f) The normalized time dependence of 670 nm PL intensity at the ring and center from (d), and 790 nm PL intensity from (e). Symbols are data points, and lines are fitting curves.

This surprising observation reveals a new perspective of the ion transport in the halide alloy: light induced ion diffusion from illuminated into non-illuminated region instead of local segregation. It is reasonable to conclude that some Br ions are expelled from the illuminated region, but there seems to exist a balancing force preventing them from either simply dispersing into the practically infinitely large area surrounding the illuminated region, i.e., forming a diffusion-length-limited decaying profile, as in the case of outward carrier diffusion under local excitation typically seen in a semiconductor [30–32]. Following the commonly accepted understanding, one would conclude that the emergence of the 790 nm emission at the illuminated region were simply due to the formation of small I-rich domains embedded in a Br-rich matrix, and nothing would happen outside the illuminated area. However, the formation of the ring exterior of the illuminated area suggests a very different scenario: the illumination generates free Br ions in the illuminated region, the concentration gradient leads to the outward diffusion of the free Br ions through halide vacancies, and the resulting electric filed due to the charge imbalance establishes a finite diffusion boundary, resembling the formation of a p-n junction in a semiconductor. Further elaboration of this idea will be provided later. Since the Br depletion region grows with time to over 4 times of the beam size in this case, it may involve an impact ion releasing process along the diffusion path, because the diffusing ions from the center have kinetic energy and momentum. Importantly, the fact that the diffusion of the Br ions into the otherwise non-perturbed area makes the PL emission there stronger suggests that the added ions passivate the halide vacancies in the ring area and leave behind more halide vacancies in the illuminated area and nearby.

To analyze the ring formation process more quantitatively, a series of line profiles corresponding to different illumination times along the white dashed lines in **Fig. 1c** are plotted in **Fig. 1 d** & **e**, respectively for the 670 and 790 nm emission. At the center, the 670 or 790 nm intensity decreases or increases monotonically with time, and at the outer edge of the ring (approximately ±40 $\mu m$), the 670 nm intensity also increases monotonically, but the intensity for a position in between exhibits non-monotonical time dependence, indicating that the intermediate region is being refilled by the Br ions from the center while in the meantime the Br ions are moving out from there. We further analyze the kinetics of the 670 nm emission intensity at the center (0 $\mu m$) and edge of the ring (±40 $\mu m$), as plotted in **Fig. 1f**, by fitting the data to appropriate exponential functions. The fitting yields time constants $\tau_c$ (670 nm) = 132 s and $\tau_r$ (670 nm) = 238 s, corresponding to the decay at the center and growth at the ring, respectively. As expected, the depletion speed is faster than that of the ring formation, because the latter is a convoluted process of the free ion generation at the illuminated area and outward diffusion to the ring. The time difference between the two processes is roughly the diffusion time. Also included in **Fig. 1f** is the time dependence of the 790 nm emission at the illumination site. The rise time of $\tau_c$ (790 nm) = 738 s is much longer than $\tau_c$ (670 nm). However, the integrated PL longer than 675 nm at the illumination site from **Fig. 1b** yields $\tau_c$ (> 675 nm) = 186 s, nearly identical to $\tau_c$ (<675 nm) = 192 s (**Fig. S1**), which indicates that the illuminated site senses the effect of Br ion being released concurrently with 670 nm emission being quenched, although some structural relaxation time, a trickling down process, is needed to reach the stable configuration that yields 790 nm emission. This finding suggests that the appearance of the 790 nm emission does not require an ion exchange process for the I ions of the exterior of the illuminated site (e.g., the ring area) to re-fill the lost Br ions, which would take a longer time since I ions are known to have smaller diffusivity [33–35]. This understanding is in stark contrast to the commonly accepted picture in which the ion segregation does not change the local charge balance. This situation is also very different from the thermal induced interdiffusion between Br-rich and I-rich regions, where a uniform alloy is achieved at the end and the process is irreversible [36,37]. Thus, the so-called "ion segregation" process is largely a redistribution of more loosely bound Br ions that might be located at the defect sites, e.g., the domain boundaries.

At the end of the study reported in **Fig. 1**, the recovery kinetics of the illuminated area is monitored in the dark (**Fig. S2**). We note that after 10 h recovery the illuminated site still exhibits about 50% stronger PL than the initial state, and the comparison of normalized spectra show nearly the same peak wavelength, but slightly less tail emission. This indicates that the recovered state of the polycrystalline material is somewhat less defective, because when the Br ions return, they might find energetically more favorable sites to form a more ordered state. The recovery process is much slower than the ring formation process, which can be explained by that the formation process has an external driving force, whereas the recovery process mainly relies on the stored energy of the electric field created by the non-equilibrium Br ion distribution. This understanding is consistent with the oscillatory behavior to be discussed later.

The phenomena reported above are universal for the beam size down to 1 $\mu m$ and up to 1 $mm$. For the largest beam size, the exterior diameter of the Br-rich ring is as large as 2 $mm$, truly macroscopic scale ion diffusion. We also note that the above findings are qualitatively valid for different illumination power densities, for instance, similar results are also observed under the 1 sun equivalent illumination, except for the process being much slower (**Fig. S3**).

To gain more insight to the chemical compositions at the center and ring region, we performed in-situ optical transmission measurements after illumination. The sample is illuminated with ~12 $\mu m$ beam at 0.3 W/cm$^2$ to generate the ring structure, with the effect shown in the inset of **Fig. 2a** for the 670 nm PL mapping. **Fig. 2a** presents the typical absorption spectra from the brightest position of the ring and the center of the illuminated region, with a reference before illumination. The absorption edge of the ring or center exhibits a moderate blue or red shift, suggesting some increase and decrease of the Br content, respectively, compared to the reference. However, the slight increase of the Br content in the ring might just result in the reduction of halide vacancies or improved crystallinity, as implied by the sharper absorption edge and reduced tail absorption. Accordingly, at the center, the red shifted absorption edge and increased tail absorption suggest that the illuminated region becomes more defective, having more halide vacancies, with slightly less Br content. The observed minor changes in the illuminated region indicates that it has not been converted into an I-rich alloy, as commonly believed, which further precludes the possibility that significant I ions move into the illuminated region while Br ions move out. Therefore, the 790 nm emission is originated by the newly created defects in the alloy rather than the I-rich alloy domains. Along this line, the generation of the mobile ions can be understood as below. The as-grown film contains a large density of defects, particularly anion vacancies. Hence, some Br ions adjacent to these defective sites are only loosely bonded to the cations in the ground state, and they are typically associated with trapped states above the valance band. Upon photoexcitation, some electrons at these defective Br sites are excited, and the bonds are further weakened, such that these ions become mobile with a concentration proportional to the power density and irradiation time. These freed ions start to diffuse away from the illuminated site via vacancies due to the concentration gradient. However, only a portion of the freed Br ions can actually diffuse away, because the recombination of the photoexcited carriers may re-anchor some disturbed Br ions. Together with the Coulomb attractive force generated by the unbalanced cations at the illuminated site, the three processes jointly establish a steady-state spatial distribution of the Br ions.

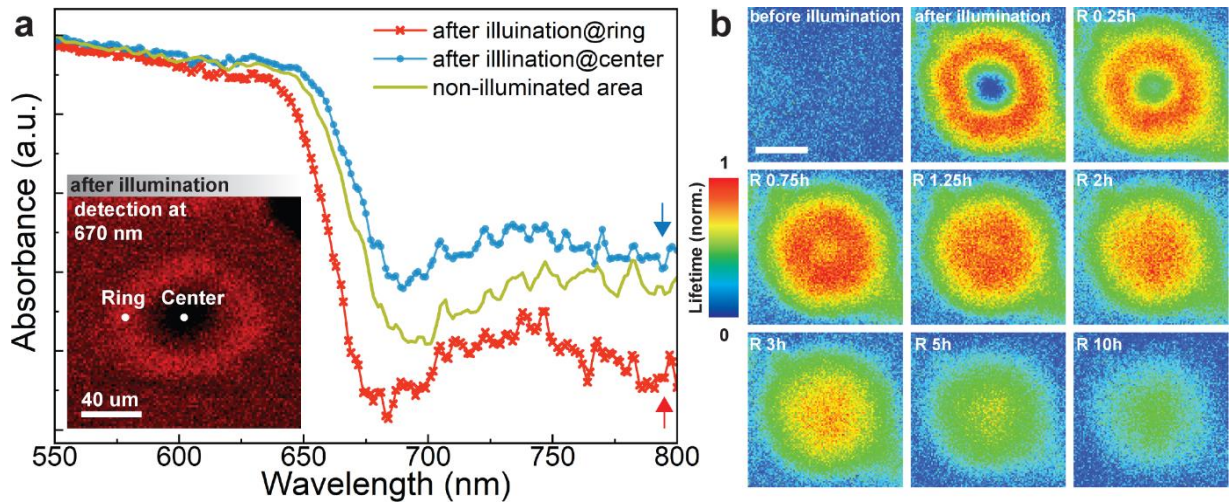

**Figure 2.** (a) Absorption spectra of the film at the brightest position of the ring and the center of the illuminated region with a reference included. Inset shows PL mapping at 670 nm after local light illumination for 1 h. (b) TRPL mapping before and after illumination and during recovery process. The scale bar is 200 $\mu m$.

TRPL may provide further insight to the ion redistribution effects observed above in the CW measurements. We perform TRPL mapping before and after local illumination and monitor the recovery. The sample is first illuminated with 130 $\mu m$ beam size and at 0.1 W/cm$^2$ for 1 h to generate the ring structure (**Fig. S3**). TRPL traces and calculated effective PL lifetimes are given in **Fig. S4**. **Fig. 2b** shows the lifetime mapping results before and after the ring formation, and at different recovery times. After illumination, at the ring area, the lifetimes are the longest and greatly increase from the before illumination value, whereas at the center, the lifetime is the shortest even below the before illumination value. However, after 10 h recovering, the lifetime at the center becomes substantially longer than the before illumination value. These findings are qualitatively consistent to the trends found in the CW results, and consistent with the ideas that after illumination the ring area exhibits better crystallinity and after recovery the center becomes less defective.

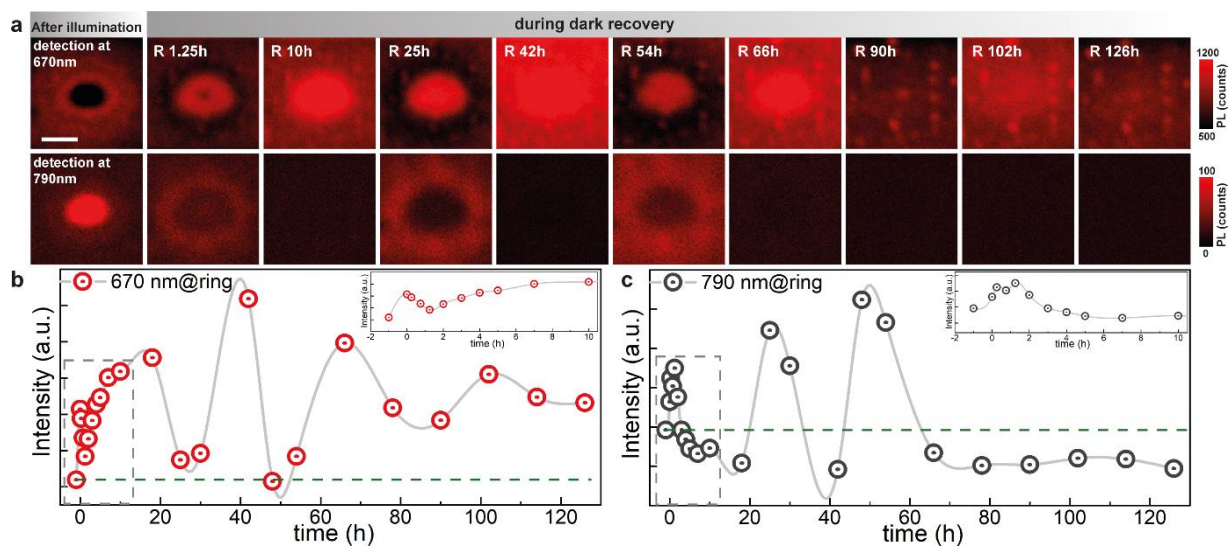

**Figure 3.** (a) PL mapping after illumination and at selected recovery times for 670 nm (top row) and 790 nm (bottom row). The scale bar is 40 $\mu m$. (b) & (c) Time dependent PL intensity taken at the ring in (a) for 670 and 790 nm, respectively. Insets show zoom-in view of the time range before 10h.

Using the measurement condition of **Fig. 1**, with lower probe power density (0.1W/cm$^2$), by monitoring longer recovery time, a striking new feature emerges: an oscillatory phenomenon is observed during recovery as shown in **Fig. 3a**. In the upper row for the 670 nm PL mapping at selected times (see **Fig. S5** for the complete series), the ring size shrinks and intensity oscillates between ring and center, while a complementary and periodic cycle appears in the 790 nm mapping. Since the "ion segregation" process is largely a redistribution of the Br ions, the oscillatory behavior suggests that the motion of the Br ions, an ionic plasma, has a restoring force that could be the Coulombic interaction between the Br ion deficit center and surplus ring. We envision a process: under a local illumination, the photo-released Br ions, like high viscosity liquid, are pushed away from the illuminated center; but, when the perturbation stops, they will return with overshooting; an oscillation will establish itself under suitable conditions. This picture is supported by the plots in **Fig. 3b** and **Fig. 3c**, the time dependent PL intensities at the ring for 670 and 790 nm. The initial intensity is at the "negative" time, whereas t = 0 h is right after illumination. The green dashed lines show the pre-illumination values. After removing the illumination, both 670 and 790 nm intensity oscillate with time, suggesting that the Br ions behavior like an ionic plasmon with a very low oscillation frequency because of the heavy ionic mass and screening. While the oscillations are eventually damped with time, they do not return to the original states before the illumination. Although we limit ourselves to a qualitative discussion here, we finish by noting that the observation of the ionic oscillation indicates that Br ions can flow and oscillate like electronic plasma. Further investigation will be required to fully understand the dynamic of the oscillation system.

In conclusion, we have provided the first demonstration of a PL ring structure, resulting from the disparity in the ionic distributions between halide anions and complementary cations in mixed halide alloys, as an analogy of the electronic based phenomenon in GaAs/AlGaAs QW systems. We have further observed an oscillatory behavior of the photo-generated free ions that could be the first reported ionic plasma oscillation in solids. On the one hand, it will open up new avenues for investigating macroscopic or mesoscopic scale ion segregation and the effects of ionic plasma in solids and applying the effects in next generation electronics (e.g., ionic patterning, self-destructive memory, energy storage); on the other hand, it has offered a completely new perspective to the halide ion segregation problem in mixed halide alloys.

Acknowledgment

XXS was supported by an ETH Zurich Postdoctoral Fellowship 19-2 EFL-55 and hosted by Professor Tiwari N. Ayodhya. His support is greatly appreciated. YZ acknowledges support of Bissell Distinguished Professorship endowment fund.

# Supplemental Material

# Photo-induced Macro/Mesoscopic Scale Ion Segregation in Mixed-Halide Perovskites: Ring Structure and Ionic Plasma Oscillations

Xiaoxiao Sun[1,2,*], Yong Zhang[3,*], Weikun Ge[4]

[1] Laboratory for Thin Films and Photovoltaics, Empa−Swiss Federal Laboratories for Materials Science and Technology, 8600 Duebendorf, Switzerland

[2] Department of Information Technology and Electrical Engineering, ETH Zurich, 8093 Zurich, Switzerland

[3] Department of Electrical and Computer Engineering, The University of North Carolina at Charlotte, Charlotte, NC 28223, USA

[4] Department of Physics, Tsinghua University, Beijing 10084, People's Republic of China

*Corresponding authors: xiasun@ethz.ch , yong.zhang@uncc.edu

# Experimental details

## Growth of perovskite films

Mixed-halide perovskite $MA_{0.17}FA_{0.83}Pb(I_{0.5}Br_{0.5})_3$ precursor solution was prepared by dissolving $PbI_2$ (0.46 M), $PbBr_2$ (0.86 M), formamidinium iodide (1 M), and methylammonium bromide (0.2 M) in a mixture of anhydrous DMF:DMSO (4:1 volume ratio, v:v). Glass substrates were cleaned by sonication in acetone and isopropyl alcohol, then the substrates were further cleaned with oxygen plasma treatment for 15 min. The perovskite solution was spin-coated on glass substrates in a two-step program at 1000 and 4000 rpm for 10 and 35 s, respectively, and 110 µL of chlorobenzene was poured on the spinning substrate 30 s after the starting of the program. The substrates were then annealed at 100 °C for 1 h. The 490±10 nm thick films were deposited in a nitrogen glove box under moisture- and oxygen- controlled conditions. After growth, the sample was broken into small pieces, which were stored in a glove box to prevent severe oxidation. Each experiment was then performed on a new piece of absorber, freshly removed from the nitrogen box. This allowed us to investigate a large number of different measurement conditions on the same sample from the same growth run.

## Spatially-resolved CW and time-resolved PL mapping

A MicroTime 100 system coupled to a detection unit from PicoQuant were used for PL/TRPL mapping. A detection fiber of 50 µm acting as a pinhole collects the light coming from the objective to the entrance of the monochromator. The image scans over an area of 10 $\mu m$ × 10 $\mu m$ to 2.5 $mm$ × 2.5 $mm$, and is composed of (100 × 100) pixels, at each pixel the dwelling time is between 5 ms to 10 ms. TRPL mapping is conducted at ~$2\times10^{12}$ photons/(cm$^2$ pulse) and 0.1 MHz frequency. For mapping at low injection, decays at each pixel were fitted to a single exponential for simplicity. PL decay curves is conducted at ~$3.2\times10^{11}$ photons/(cm$^2$ pulse) and 0.1 MHz frequency at illuminated area and ring.

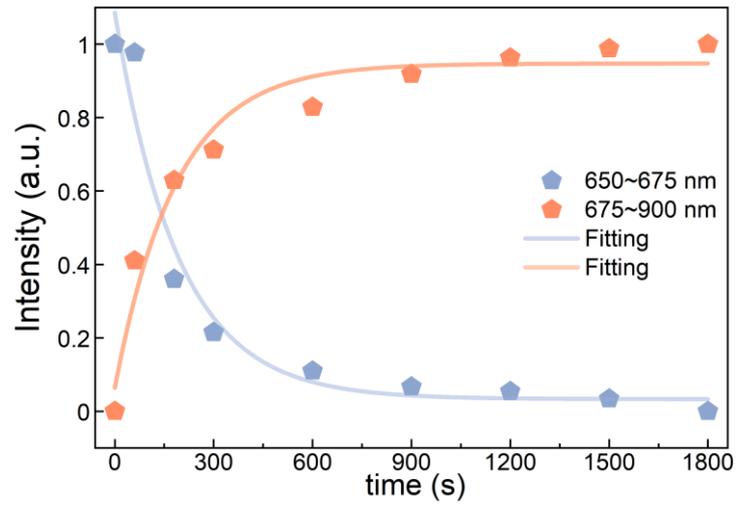

**Figure S1.** The normalized integrated intensity below 675 nm and above 675 nm over time at the illumination site from Figure 1b.

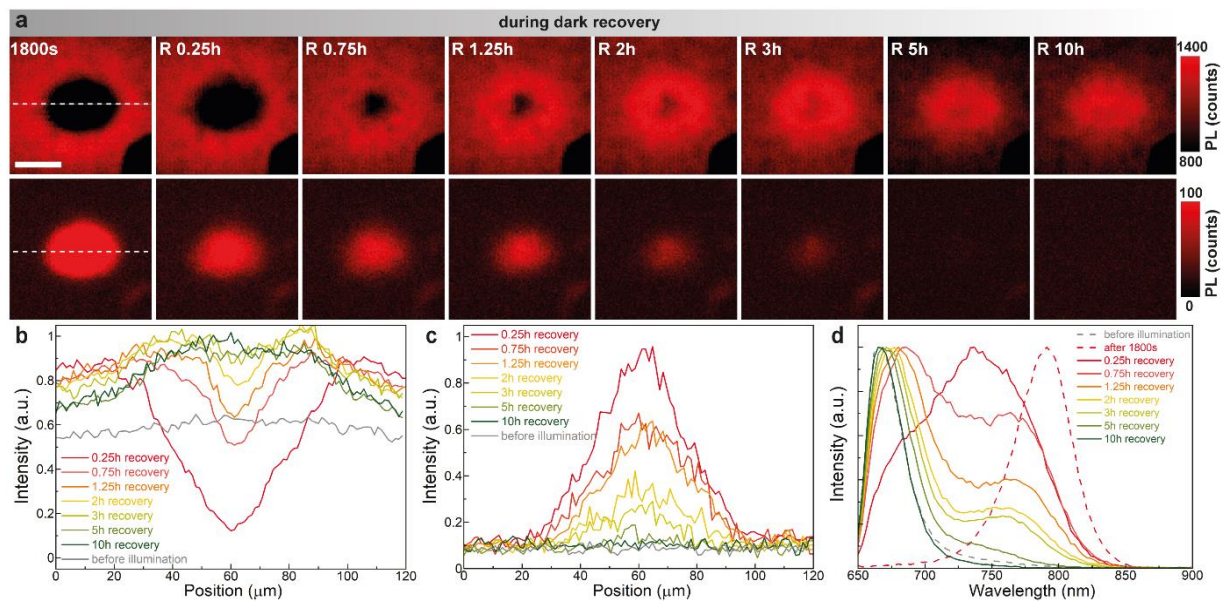

**Figure S2.** Recovery process of light induced ion redistribution. (a) PL mapping at different recovery times for 670 nm (top row) and 790 nm (bottom row). The scale bar is 40 $\mu m$. (b) & (c) Time dependent PL spatial profiles taken at the white dashed lines in (a) for 670 and at 790 nm, respectively. (d) Dynamic tracking of normalized PL spectra at the illuminated site.

In the upper row for 670 nm, the size of the ring shrinks gradually and the ring disappears after about 3 hrs. However, the illuminated site does not recover the initial state but exhibits about 50% stronger PL than the initial state, and remains so even after 10 hrs. The lower row shows the corresponding results for the 790 nm emission. The bright region contracts with time and the PL intensity recovers the initial state in about 5 hrs. The changes in intensity can be better seen in **Fig. S2b** and **c** for the line profiles along the dashed lines in **Fig. S2a**, respectively, for 670 and 790 nm, whereas the before illumination profiles are also included for comparison. During the recovery process, we also collect the PL spectra at the center, as shown in **Fig. S2d**. The comparison of normalized spectra show nearly the same peak wavelength, but slightly less tail emission.

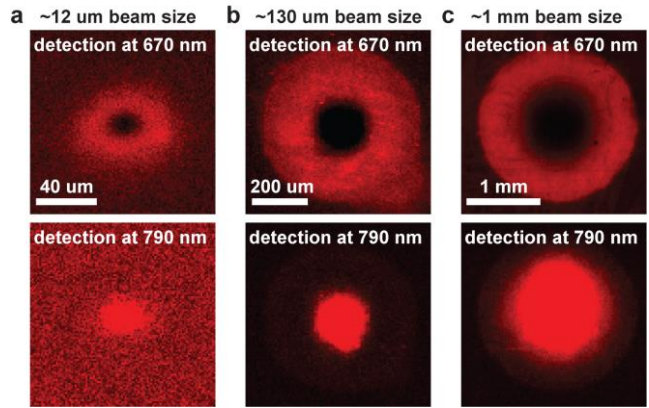

**Figure S3.** PL emission mapping at detection wavelength of 670 nm and 790 nm after equivalent 1 sun illumination for 1 h at beam size of (a) ~12 um, (b) ~130 um and (c) ~ 1 mm.

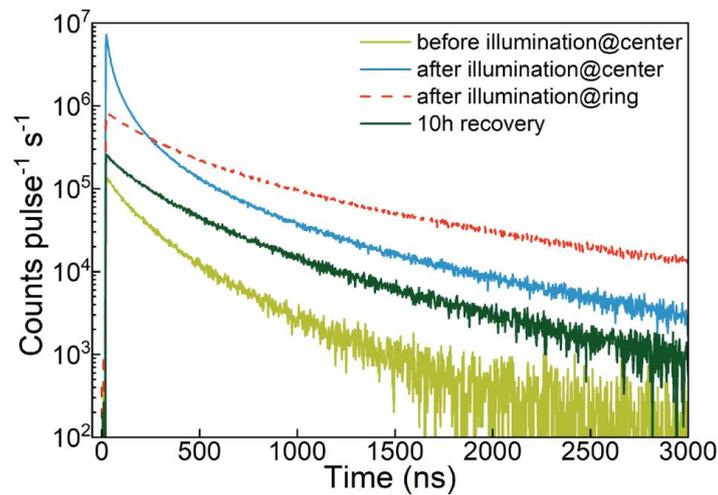

**Figure S4.** The typical PL decay traces: before illumination, from the center and ring at the end of the illumination, and from the center after 10 h recovery. The PL decay traces can be fitted well using a biexponential function $I(t) = A_1 exp(-t/\tau_1) + A_2 exp(-t/\tau_2)$, and an effective decay lifetime can be defined as $\tau_{avg} = \frac{A_1\tau_1^2 + A_2\tau_2^2}{A_1\tau_1 + A_2\tau_2}$.

| Time | $A_1$ [Cnts] | $\tau_1$ [ns] | $A_2$ [Cnts] | $\tau_2$ [ns] | $\frac{\sum_i A_i \cdot \tau_i^2}{\sum_i A_i \cdot \tau_i}$ |
|---|---|---|---|---|---|
| Before illumination | 110714.4 | 100.9365 | 50669.44 | 341.079 | 246.7759 |
| After at center | 13766072 | 24.12772 | 1873947 | 173.9286 | 98.32126 |
| After at ring | 649656.7 | 240.1195 | 225465.4 | 1024.277 | 708.1386 |
| R 10h at center | 181235.6 | 145.9546 | 99207.18 | 533.1424 | 404.059 |

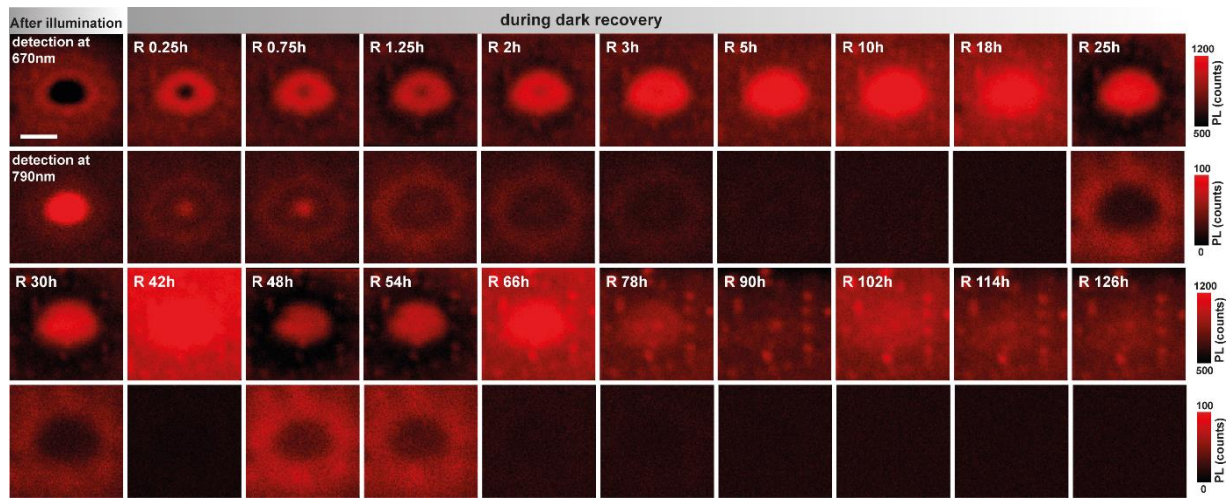

**Figure S5.** PL mapping after illumination and during recovery for 670 nm and 790 nm.